\documentclass[11pt,a4paper]{article}
\usepackage{jheppub}

\usepackage[latin1]{inputenc} 
\usepackage[T1]{fontenc}
\usepackage[safe]{textcomp}
\usepackage{lmodern} 
\usepackage{epsfig}
\usepackage{float}
\usepackage{amstext}  
\usepackage{amsfonts}
\usepackage{amssymb} 
\usepackage{amsmath}
\usepackage{amsthm}
\usepackage{bm}       
\usepackage[bottom]{footmisc} 
\usepackage{array}                
\usepackage{xcolor} 
\usepackage{framed}
\usepackage{slashed}
\usepackage[absolute]{textpos}
\usepackage{axodraw4j}
\usepackage{dsfont}
\usepackage{multirow}

\usepackage{appendix}
\usepackage{listings}
\usepackage{enumerate}     
\usepackage[bottom]{footmisc} 
\usepackage{array}           
\usepackage{parskip}        
\usepackage{xcolor} 
\usepackage{color}
\usepackage{framed}
\usepackage{fancyhdr}
\usepackage{subfig}

\usepackage{tikz}
\usepackage[absolute]{textpos}
\usepackage{pstricks}
\usepackage{axodraw4j}
\usepackage{graphicx}

\newcommand{\p}{\partial}

\newcommand{\f}[2]{\frac{#1}{#2}}
\newcommand{\sss}[1]{\scriptscriptstyle{#1}}
\newcommand{\ssst}[1]{\scriptscriptstyle{\text{#1}}}

\newcommand{\bea}{\begin{eqnarray}}
\newcommand{\eea}{\end{eqnarray}}
\newcommand{\be}{\begin{equation}}
\newcommand{\ee}{\end{equation}}
\newcommand{\ba}{\begin{align}}
\newcommand{\ea}{\end{align}}
\newcommand{\beas}{\begin{eqnarray*}}
\newcommand{\eeas}{\end{eqnarray*}}
\newcommand{\bes}{\begin{equation*}}
\newcommand{\ees}{\end{equation*}}
\newcommand{\bas}{\begin{align*}}
\newcommand{\eas}{\end{align*}}

\newcommand{\ssL}{{\mathcal L}} 
\newcommand{\eps}{{\varepsilon}}
 
\newcommand{\cf}{C_{\scriptscriptstyle{F}}} 
\newcommand{\ca}{C_{\scriptscriptstyle{A}}}
\newcommand{\tr}{T_{\scriptscriptstyle{F}}}
\newcommand{\dF}{d_{\scriptscriptstyle{F}}}

\newcommand{\Ng}{N_{\scriptscriptstyle{A}}}

\newcommand{\Ngl}{n_{\scriptscriptstyle{\tilde{g}}}}

\newcommand{\Nf}{n_{\scriptscriptstyle{f}}}
\newcommand{\gs}{g_{\scriptscriptstyle{s}}}
\newcommand{\yt}{y_{\scriptscriptstyle{t}}}

\newcommand{\als}{\alpha_{\scriptscriptstyle{s}}}

\newcommand{\lb}{\left(}
\newcommand{\rb}{\right)}

\newcommand{\msbar}{$\overline{\text{MS}}$}

\newcommand{\dAAfNg}{\frac{d_{\scriptscriptstyle{A}}^{abcd}d_{\scriptscriptstyle{A}}^{abcd}}{\Ng}}

\newcommand{\dQQfNg}{\frac{d_{\scriptscriptstyle{F}}^{abcd}d_{\scriptscriptstyle{F}}^{abcd}}{\Ng}}
\newcommand{\dQAfNg}{\frac{d_{\scriptscriptstyle{F}}^{abcd}d_{\scriptscriptstyle{A}}^{abcd}}{\Ng}}

\newcommand{\dFFfijNg}{\frac{d_{\scriptscriptstyle{F,i}}^{abcd}d_{\scriptscriptstyle{F,j}}^{abcd}}{\Ng}}
\newcommand{\dFAfiNg}{\frac{d_{\scriptscriptstyle{F,i}}^{abcd}d_{\scriptscriptstyle{A}}^{abcd}}{\Ng}}
\newcommand{\cfi}{C_{\scriptscriptstyle{F,i}}} 
\newcommand{\cfj}{C_{\scriptscriptstyle{F,j}}} 
 
\newcommand{\tri}{T_{\scriptscriptstyle{F,i}}}
\newcommand{\dFi}{d_{\scriptscriptstyle{F,i}}}
\newcommand{\trj}{T_{\scriptscriptstyle{F,j}}}

\newcommand{\trk}{T_{\scriptscriptstyle{F,k}}}

\newcommand{\Nfi}{n_{\scriptscriptstyle{f,i}}}
\newcommand{\Nfj}{n_{\scriptscriptstyle{f,j}}}
\newcommand{\Nfk}{n_{\scriptscriptstyle{f,k}}}

\definecolor{bluemar}{rgb}{0,0,.5}
\definecolor{redmar}{rgb}{.8,0,0}
\definecolor{greenmar}{rgb}{0,.5,0}

\newcommand{\bigrint}{%
\hbox to 2.92em{\hss\scalebox{1.1}[1] {\rotatebox[origin=c]{15}{$\displaystyle\int$}}\hss}}

\catcode`\@=11
\def\slash{\mathpalette\make@slash}
\def\make@slash#1#2{\setbox\z@\hbox{$#1#2$}%
  \hbox to 0pt{\hss$#1/$\hss\kern-\wd0}\box0}
\catcode`\@=12 

\newcommand{\beq}{\begin{equation}}
\newcommand{\eeq}{\end{equation}}

\allowdisplaybreaks
\newcommand{\ice}[1]{\relax}

\title{Four-loop QCD $\beta$-function with different fermion representations of the gauge group}
\author[a]{M.~F.~Zoller}
  \affiliation[a]{Institut f\"ur Physik, University of Zurich (UZH), Switzerland}

\emailAdd{zoller@physik.uzh.ch}

\abstract{We present analytical results at four-loop level for the $\beta$-function of the coupling of a generic gauge group 
and any number of different quark representations. From this we can directly derive the gluino contribution to the strong coupling $\beta$-function of
supersymmetric extensions of the Standard Model.}

\arxivnumber{}
\subheader{ZU-TH-32/16}

\begin{document}
\maketitle

\section{Introduction}

The Renormalization Group (RG) functions of non-Abelian gauge theories, especially the QCD $\beta$-function, are among the most precisely 
calculated objects in quantum field theory. The excellent agreement of theory predictions for the strong coupling at different scales with experimental 
results is among the great successes of particle physics, since the observed asymptotic freedom is the basis for theory predictions at hadron 
colliders. 

An interesting special case in theoretical physics are conformal theories in which the $\beta$-functions vanish and the couplings are
hence constant. The knowledge of the QCD $\beta$--function with an extended fermion sector is an important ingredient for the application of the
sequential extended BLM approach \cite{Mikhailov:2004iq,Kataev:2014zwa,Kataev:2014jba}, which aims at resumming the non-conformal
parts of QCD observables into the scale of the coupling in a unique way, to extensions of the Standard Model (SM). A closely related approach is the Principle of 
Maximum Conformality and Commensurate Scale Relations developed in \cite{Brodsky:2013vpa}.

The $\beta$-function for the coupling $\als=\f{\gs^2}{4\pi}$ is defined as
\be
\beta(\als)=\mu^2\f{d \als}{d \mu^2}    =\als\sum \limits_{n=1}^{\infty} \left(\f{\als}{4\pi}\right)^{n}\,\beta_{\als}^{\sss{(n)}}{}.
\ee
and has been computed at one-loop \cite{PhysRevLett.30.1343,PhysRevLett.30.1346}, two-loop \cite{Jones1974531,Tarasov:1976ef,PhysRevLett.33.244,
Egorian:1978zx}, three-loop \cite{Tarasov:1980au,3loopbetaqcd} and four-loop \cite{4loopbetaqcd,Czakon:2004bu} 
level for a generic gauge group with one fermion representation.\footnote{The RG functions of the full Standard Model 
are available at three-loop order for the gauge couplings \cite{PhysRevLett.108.151602,Mihaila:2012pz,Bednyakov:2012rb}, the Yukawa couplings \cite{Chetyrkin:2012rz,Bednyakov:2012en}
and the parameters of the Higgs potential \cite{Chetyrkin:2012rz,Chetyrkin:2013wya,Bednyakov:2013eba,Bednyakov:2013cpa}.
The four-loop $\beta$-function for the strong coupling $\gs$ was extended to include the dependence on the top-Yukawa 
coupling $\yt$ and the Higgs self-coupling $\lambda$ \cite{Bednyakov:2015ooa,Zoller:2015tha}. 
The leading QCD induced four-loop contributions to the Higgs self-coupling $\beta$-function were
presented in \cite{Martin:2015eia,Chetyrkin:2016ruf}.} 
Recently the five-loop result was published for QCD colour factors
\cite{Baikov:2016tgj} and the terms $\propto \Nf^3$ and $\propto \Nf^4$ for a generic gauge group \cite{Luthe:2016ima}.

In this paper we present the four-loop $\beta$-function for the coupling of a generic gauge group with an arbitrary number $N_{\ssst{rep}}$
of fermion representations. One example is the extension of the QCD $\beta$-function to include
not only $\Nf$ active quark flavours but also $\Ngl$ gluinos. At two-loop \cite{Jones:1974pg} and three-loop \cite{Clavelli:1996pz} level these results are
available and are independently confirmed here. Recently, also the four-loop result for the gluino case was presented 
at a conference \cite{BednyakovTalkBetags4l}.

\section{Details of the calculation}

\subsection{QCD with several fermion representations}

The QCD Lagrangian is given by
\bea
\ssL_{\sss{QCD}}&=&-\f{1}{4}G^a_{\mu \nu} G^{a\,\mu \nu}-\f{1}{2 (1-\xi)}\lb\p_\mu A^{a\,\mu}\rb^2 
+\p_\mu \bar{c}^a \p^{\mu}c^a+\gs f^{abc}\,\p_\mu \bar{c}^a A^{b\,\mu} c^c \nonumber \\
&+&\sum\limits_q\left\{\f{i}{2}\bar{q}\overleftrightarrow{\slashed{\p}}q+ \gs \bar{q}\slashed{A}^a T^a q\right\}{},
\label{LQCD} 
\eea  
with the gluon field strength tensor
\be
G^a_{\mu \nu}=\p_\mu A^a_\nu - \p_\nu A^a_\mu + \gs f^{abc}A^b_\mu A^c_\nu{},
\ee
the structure constants $f^{abc}$ of the gauge group, defined through
\be \left[ T^a,T^b \right]=if^{abc}T^c\ee
with the group generators $T^a$, and $q$ running over all quark flavours.

This can easiliy be generalized to include several fermion representations of the gauge group by substituting
\be 
\sum\limits_q\left\{\f{i}{2}\bar{q}\overleftrightarrow{\slashed{\p}}q+ \gs \bar{q}\slashed{A}^a T^a q\right\}
\to
\sum\limits_{r=1}^{N_{\ssst{rep}}}\sum\limits_{q_r}
\left\{\f{i}{2}\bar{q_r}\overleftrightarrow{\slashed{\p}}q_r+ \gs \bar{q_r}\slashed{A}^a T_r^a q_r\right\}
\ee
where $r$ gives the representation and $q_r$ runs over all fermion flavours/types of that representation.

For example in the strong sector of supersymmetric extensions of the Standard Model (SM) we find in addition to $\Nf$ quark flavours, which are Dirac fermions,
 $\Ngl$ gluinos, which are Majorana fermions. In the SM we have  $\Ngl=0$, in the MSSM  $\Ngl=1$.

This extended QCD Lagrangian is renormalized with the counterterm Lagrangian
\bea
\delta\!\ssL_{\sss{QCD}}&=&-\f{1}{4}\delta\! Z^{(2g)}_3 \lb \p_\mu A^a_\nu - \p_\nu A^a_\mu \rb^2
-\f{1}{2}\delta\! Z^{(3g)}_1 \gs f^{abc}\lb \p_\mu A^a_\nu - \p_\nu A^a_\mu \rb A^b_\mu A^c_\nu \nonumber\\
&-&\f{1}{4}\delta\! Z^{(4g)}_1 \gs^2 \lb f^{abc} A^b_\mu A^c_\nu \rb^2
+\delta\! Z^{(2c)}_3 \p_\mu \bar{c}^a \p^{\mu}c^a+\delta\! Z^{(ccg)}_1 \gs f^{abc}\,\p_\mu \bar{c}^a A^{b\,\mu} c^c \\
&+& \sum\limits_{r=1}^{N_{\ssst{rep}}}\sum\limits_{q_r}
\left\{Z^{(q_rq_r)}_2\f{i}{2}\bar{q_r}\overleftrightarrow{\slashed{\p}}q_r+ \gs Z^{(q_rq_rg)}_1\bar{q_r}\slashed{A}^a T_r^a q_r\right\}
\label{LQCDc}
\eea
The renormalization constant for the strong coupling can then be computed from
\be 
Z_{\gs}=\f{Z^{(ccg)}_{1}}{Z^{(2c)}_{3}\sqrt{Z^{(2g)}_{3}}} \label{Zgscomputation}
\ee
where we define the renormalization constants $Z=1+\delta Z$ in the \msbar-scheme. 
All divergent integrals are regularized in $D=4-2 \eps$ space time dimensions.

\subsection{Technicalities}

All 1-particle-irreducible Feynman diagrams for the computation of $Z^{(2c)}_3$, $Z^{(2g)}_3$ and $Z^{(ccg)}_1$ are generated with QGRAF \cite{QGRAF}. 
The C++ programs Q2E and EXP \cite{Seidensticker:1999bb,Harlander:1997zb} identify the topology of the diagram. 
The UV divergent part of the diagrams is calculated as described in detail in \cite{Zoller:2014xoa}
introducing the same auxiliary mass parameter $M^2$ in every propagator denominator and cancelling subdivergencies $\propto M^2$
by introducing an unphysical gluon mass counterterm \mbox{$\f{M^2}{2}\delta\!Z_{\sss{M^2}}^{(2g)}\,A_\mu^a A^{a\,\mu}$}. 
This procedure was introduced in \cite{Misiak:1994zw}, further developed in \cite{beta_den_comp} and also used e.~g.~in \cite{4loopbetaqcd,Czakon:2004bu,Chetyrkin:2012rz,
Zoller:2015tha,Chetyrkin:2016ruf}. For a detailed explanation of this method see e.~g.~\cite{Zoller:2014xoa}.

In the \msbar{}-scheme renormalization constants are independent of external momenta.
Hence we Taylor expand to first order in the external momentum $q$ for the ghost-gluon vertex\footnote{$q$ being the external momentum
entering the ghost leg, the external momentum of the gluon is set to zero from the start. $\mu$ is the Lorentz index of the gluon.}
(which is $\propto q^{\mu}$) and to second order in the gluon self-energy ($\propto q^{\mu}q^{\nu}-q^2 g^{\mu\nu}$) as well as 
the ghost self-energy ($\propto q^2$).
Then we use projectors on the integrals in order to make them scalar and $q$-independent, namely $\f{q^\mu}{q^2}$ 
for the ghost-gluon vertex, $\f{1}{q^2}$ for the ghost propagator and $\f{q^\mu q^\nu}{q^4}$ as well as $\f{g^{\mu\nu}}{q^2}$
for the gluon propagator. 
These expansions and projections as well as the fermion traces and counterterm insertions in lower order diagrams were
done with FORM \cite{Vermaseren:2000nd,Tentyukov:2007mu}. The resulting tadpole integrals were computed with
the \mbox{FORM}-based package \mbox{MATAD} \cite{MATAD} up to three-loop order. At four loops we use the C++ version 
of FIRE 5 \cite{Smirnov:2008iw,Smirnov:2014hma} in order to reduce the scalar integrals
to Master Intgrals which are available from \cite{Czakon:2004bu}.
For details on the reduction we refer to the previous paper \cite{Zoller:2015tha}.

The computation of the gauge group factors is based on the \mbox{FORM} package COLOR \cite{COLOR}.
The implementation by the authors of \cite{COLOR} takes as input the colour diagrams expressed
through the generators of the fermion representation $T^a_{ij}=\texttt{T(i,j,a)}$ and the structure constants 
$f^{abc}=\texttt{f(a,b,c)}$, i.~e.~the 
generators of the adjoint representation of the corresponding Lie algebra. The result is then given in terms of Casimir operators $\cf$ and $\ca$
of the fermion and adjoint representation, the trace $\tr$ defined through 
\be \tr \delta^{ab}=\textbf{Tr}\lb T^{a} T^{b}\rb=T^{a}_{ij} T^{b}_{ji} \ee
and higher order invariants built from symmetric tensors 
\be d_{\sss{R}}^{a_1 a_2 \ldots a_n}=\f{1}{n!} 
\sum\limits_{\text{perm } \pi}\text{Tr}\left\{ T^{a_{\pi(1)},R}T^{a_{\pi(2)},R}\ldots T^{a_{\pi(n)},R}\right\}{},\ee
where $R$ marks the representation, in this case $T^{a,F}_{i,j}=T^{a}_{i,j}$ or $T^{a,A}_{bc}=-i\,f^{abc}$.
The fermion representation has dimension $\dF$ and the adjoint $\Ng$. Each fermion loop is multiplied by the number of
active fermion flavours $\Nf$.

We extend this setup to take as input generators $T^{a,1}_{ij}=\texttt{T1(i,j,a)}$, 
$T^{a,2}_{ij}=\texttt{T2(i,j,a)}$, $T^{a,3}_{ij}=\texttt{T3(i,j,a)}$, $\ldots$ in order to account for different representations. The adjoint representation
with the structure constants as generators is always present. The traces of these generators which are the objects
simplified and reduced by COLOR also carry an index for the representation and all internal routines of the COLOR package
as described in \cite{COLOR} are adapted to keep track of the fermion representation to which the produced invariants
belong. We therefore have Casimir operators $\cfi$ for the fermion
representations ($i=1,\ldots,N_{\ssst{rep}}$) of dimensions $\dFi$. The trace for each representation is given by $\tri$
defined as \mbox{$\tri \delta^{ab}=\textbf{Tr}\lb T^{a,i} T^{b,i}\rb$} and the higher order invariants are constructed
from \bea
d_{\scriptscriptstyle{F,i}}^{abcd} &=& \f{1}{6} \text{Tr} 
\lb T^{a,i} T^{b,i} T^{c,i} T^{d,i} + T^{a,i} T^{b,i} T^{d,i} T^{c,i} + T^{a,i} T^{c,i} T^{b,i} T^{d,i} \right. \nonumber \\
&+& \left. T^{a,i} T^{c,i} T^{d,i} T^{b,i} + T^{a,i} T^{d,i} T^{b,i} T^{c,i} + T^{a,i} T^{d,i} T^{c,i} T^{b,i} \rb{}.
\eea
from the generators of the fermion representations and analogously $d_{\scriptscriptstyle{A}}^{abcd}$ constructed from the 
generators of the adjoint representation. A closed fermion loop is accompanied by a factor $\Nfi$ for the number of active
fermion flavours of representation $i$.

Now, we could take a model with a fixed number of fermion representations, each given by its own field in QGRAF,
and perform the calculation directly. But there is a more elegant way: 
We only use one field $q$ in QGRAF for all fermion representations. Each fermion line then gets a different 
``representation'' number $R=1,2,3$ in our modified version of COLOR. However, this index $R$ is just the number of the fermion line and
does not correspond to one single representation of the gauge group, but in the end runs over all representations.

A diagram with one fermion line will then yield a result involving the trace \texttt{TF1}
and the Casimirs \texttt{cF1} and \texttt{cA} as well as the dimensions \texttt{dF1} and \texttt{NA}.
A diagram with two fermion lines results in \texttt{TF1}, \texttt{TF2}, \texttt{cA}, \texttt{dF1}, \texttt{dF2} and 
\texttt{NA}. In our four-loop calculation we encounter at most three fermion lines in the gauge boson and ghost self-energies
and the ghost-gauge boson vertex correction.
In the end we sum each index over all physical fermion representations because each fermion loop can be any fermion type. 
The advantage is obviously that we do not generate more diagrams than in pure QCD! In the same way that we sum over all quark types in QCD 
(using only one quark field in QGRAF and multiplying quark loops with $\Nf$) we now sum over all fermion types, 
e.~g.~all quarks and all gluinos in the MSSM. 

For the renormalization procedure, however, it is convenient to deal with explicit group factors.
Since we can encounter at most three physical representations in one and the same diagram
we substitute, e.~g.~in a one-loop diagram
\be \texttt{Nf*TF1} \to n_{\sss{f,1}} T_{\sss{F,1}} + n_{\sss{f,2}} T_{\sss{F,2}} + n_{\sss{f,3}} T_{\sss{F,3}}. \ee
where the $1$ on the lhs marks the number of the fermion line and the $1,2,3$ on the rhs correspond to different
representations of the gauge group.\footnote{The substiution rules can become quite involved in higher order diagrams. It
is therefore convenient to collect all combinations $\texttt{Nf}^\texttt{x1}\texttt{*TF1}^\texttt{x2}\texttt{*CF1}^\texttt{x3}
\texttt{*TF2}^\texttt{x4}\texttt{*CF2}^\texttt{x5}\texttt{*TF3}^\texttt{x6}\texttt{*CF3}^\texttt{x7}$
in a function \texttt{C(x1,\ldots,x7)}. The factors \texttt{C(x1,\ldots,x7)} are then substituted by the proper 
symmetrization over three representations.}
In this way we compute the gauge boson, ghost and fermion self-energy as well as the ghost-gauge boson and 
fermion-gauge boson vertex corrections at one-loop, two-loop and three-loop level in order to have the correct 
counterterms available.
The explicit use of all fermion representations 
in the counterterms as well as in the diagrams in which they are inserted and the diagrams to which the
lower-loop diagrams with counterterms are added is necessary
in order to ensure that e.~g.~the one-loop counterterm $\delta Z_3^{(2g)}$ can be used to cancel the subdivergence
stemming from the first as well as from the second fermion loop in a higher-loop diagram. 

After all lower-loop diagrams with counterterms are added to the diagrams of the loop order we want to compute
we can simplify the notation again by substituting \footnote{Again it is useful to collect 
$n_{\sss{f,1}}^{x_1} n_{\sss{f,2}}^{x_2} n_{\sss{f,3}}^{x_3}
 T_{\sss{F,1}}^{x_4} T_{\sss{F,2}}^{x_5} T_{\sss{F,3}}^{x_6} 
 C_{\sss{F,1}}^{x_7} C_{\sss{F,2}}^{x_8} C_{\sss{F,3}}^{x_9}$ in a function
\texttt{CR(x1,\ldots,x9)} for which then the substitution rules are formulated. This also provides a check
that all terms have indeed been absorbed into the compact notation when no \texttt{CR} survives in the end.}
, e.~g.~
\be n_{\sss{f,1}} T_{\sss{F,1}} \to \sum\Nfi\tri - n_{\sss{f,2}} T_{\sss{F,2}} - n_{\sss{f,3}} T_{\sss{F,3}}. \ee
Note that the final result which only contains $\Nfi, \tri, \cfi, \dFi,\ldots$ is not restricted to three representations,
but is valid for any number $N_{\sss{rep}}$. 

\section{Results \label{res:beta}}
In this section we give the results for the four-loop $\beta$-function of the strong coupling $\gs$ with an arbitrary number of fermion representations
The number of active fermion flavours of representation $i$ is denoted by $\Nfi$.
\bea
\beta_{\sss{\als}}^{(1)}&=& 
          - \f{11}{3} \ca 
          +\sum\limits_{i} \f{4}{3} \Nfi \tri   \label{1lbetaas}        \\
\beta_{\sss{\als}}^{(2)}&=&           - \f{34}{3} \ca^2
          +\sum\limits_{i} \Nfi \tri \left[           4 \cfi
          + \f{20}{3} \ca \right]\label{2lbetaas}\\
\beta_{\sss{\als}}^{(3)}&=&           - \f{2857}{54} \ca^3 
          + \sum\limits_{i} \Nfi \tri \left[
          - 2 \cfi^2
          + \f{205}{9} \ca \cfi
          + \f{1415}{27} \ca^2 \right]  \nonumber\\ &-& 
          \sum\limits_{i,j} \Nfi \Nfj \tri\trj\left[
           \f{44}{9} \cfi           
          + \f{158}{27} \ca \right]\label{3lbetaas}\\
\beta_{\sss{\als}}^{(4)}&=&
          -\lb  \f{150653}{486} - \f{44}{9} \zeta_{3} \rb \ca^4
           + \lb \f{80}{9} - \f{704}{3} \zeta_{3} \rb \dAAfNg \nonumber\\
          &+& \sum\limits_{i} \Nfi \tri \left[- 46 \cfi^3 
          + \lb \f{4204}{27}   - \f{352}{9} \zeta_{3} \rb \ca \cfi^2  
          - \lb \f{7073}{243}  - \f{656}{9} \zeta_{3} \rb \ca^2 \cfi \right.\nonumber\\ & & \left.
          + \lb \f{39143}{81}  - \f{136}{3} \zeta_{3} \rb \ca^3  \right] 
          - \sum\limits_{i} \Nfi \lb \f{512}{9} - \f{1664}{3} \zeta_{3} \rb \dFAfiNg  \label{4lbetaas}\\
          &+& \sum\limits_{i,j} \Nfi \Nfj\tri \trj\left[
          - \lb \f{184}{3} - 64 \zeta_{3} \rb \cfi  \cfj
          + \lb \f{304}{27}  + \f{128}{9} \zeta_{3} \rb \cfi^2  \right.\nonumber\\ & & \left.
          - \lb \f{17152}{243}  + \f{448}{9} \zeta_{3} \rb \ca  \cfi 
          - \lb \f{7930}{81} + \f{224}{9} \zeta_{3} \rb \ca^2  \right]\nonumber\\ 
          &+& \sum\limits_{i,j}\Nfi \Nfj \lb \f{704}{9} 
          - \f{512}{3}  \zeta_{3} \rb \dFFfijNg \nonumber\\
          &-& \sum\limits_{i,j,k} \Nfi \Nfj  \Nfk\tri\trj  \trk\left[ \f{1232}{243} \cfi 
          + \f{424}{243} \ca \right]\nonumber
\eea
The pure QCD part of \eqref{4lbetaas} agrees with \cite{4loopbetaqcd,Czakon:2004bu}.

The special case of QCD plus additional gluinos is derived by choosing $N_{\sss{rep}}=2$ and
\be 
\parbox{0.8\textwidth}{
\begin{tabular}{llllll}
$n_{\scriptscriptstyle{f,1}}$ & $=$ & $\Nf,\qquad$ & $n_{\scriptscriptstyle{f,2}}$ & $=$ & $\f{\Ngl}{2},$\\
$T_{\scriptscriptstyle{F,1}}$ & $=$ & $\tr,\qquad$ & $T_{\scriptscriptstyle{F,2}}$ & $=$ & $\ca,$\\
$C_{\scriptscriptstyle{F,1}}$ & $=$ & $\cf,\qquad$ & $C_{\scriptscriptstyle{F,2}}$ & $=$ & $\ca,$
\end{tabular}
}
\ee
where the factor $\f{1}{2}$ in $n_{\scriptscriptstyle{f,2}}=\f{\Ngl}{2}$ is due to the Majorana nature of the gluinos (see e.~g.~\cite{Clavelli:1996pz}). We find
\bea
\beta_{\sss{\als}}^{(1)}&=&  - \f{11}{3} \ca
          + \f{2}{3} \Ngl \ca
          + \f{4}{3} \Nf \tr
\label{1lbetaasgl}        \\
\beta_{\sss{\als}}^{(2)}&=&    - \f{34}{3} \ca^2
          + \f{16}{3} \Ngl \ca^2 
          + \Nf \tr \left[ 4 \cf + \f{20}{3} \ca \right]
\label{2lbetaasgl}\\
\beta_{\sss{\als}}^{(3)}&=&    - \f{2857}{54} \ca^3
          + \f{988}{27} \Ngl \ca^3          
          - \f{145}{54} \Ngl^2 \ca^3 
          +\Nf \tr \left[          - 2 \cf^2          + \f{205}{9} \ca \cf          + \f{1415}{27} \ca^2 \right]\nonumber\\ & &
          -\Nf^2 \tr^2 \left[            \f{44}{9} \cf          + \f{158}{27}  \ca \right]
          -\Ngl \Nf \ca \tr \left[ \f{22}{9} \cf  + \f{224}{27} \ca \right]
\label{3lbetaasgl}\\
\beta_{\sss{\als}}^{(4)}&=&
          - \lb \f{150653}{486} - \f{44}{9} \zeta_{3}\rb \ca^4          + \lb\f{80}{9} - \f{704}{3} \zeta_{3} \rb\dAAfNg \nonumber\\ & &
          +\Ngl\left[ \lb \f{68507}{243} - \f{52}{9} \zeta_{3} \rb\ca^4 - \lb\f{256}{9} - \f{832}{3} \zeta_{3}\rb\dAAfNg \right]  \nonumber\\ & &
          -\Ngl^2\left[\lb \f{26555}{486} - \f{8}{9} \zeta_{3} \rb\ca^4 + \lb\f{176}{9} - \f{128}{3} \zeta_{3}\rb\dAAfNg \right]  
          - \f{23}{27} \Ngl^3 \ca^4 \nonumber\\ & &
          + \Nf \tr \left[
          - 46  \cf^3
          + \lb\f{4204}{27} - \f{352}{9} \zeta_{3}\rb \ca \cf^2
          - \lb\f{7073}{243} - \f{656}{9} \zeta_{3}\rb\ca^2 \cf \right. \nonumber\\ & &\left.
          + \lb\f{39143}{81} - \f{136}{3} \zeta_{3}\rb \ca^3 
          \right]
          -  \Nf \lb\f{512}{9} - \f{1664}{3}\zeta_{3}\rb \dQAfNg  \nonumber\\ & &
          + \Nf^2 \tr^2 \left[
          - \lb\f{1352}{27} - \f{704}{9} \zeta_{3}  \rb\cf^2
          - \lb\f{17152}{243} + \f{448}{9} \zeta_{3} \rb\ca  \cf \right. \nonumber\\ & &\left.
          - \lb\f{7930}{81}  + \f{224}{9} \zeta_{3} \rb\ca^2 
          \right]
          + \Nf^2 \lb\f{704}{9}  - \f{512}{3} \zeta_{3}\rb \dQQfNg  \nonumber\\ & &
          -\Nf^3 \tr^3 \left[
           \f{1232}{243} \cf
          + \f{424}{243}  \ca 
          \right] \label{4lbetaasgl} \\ & &
          +\Ngl \Nf \ca \tr \left[
           \lb\f{152}{27} + \f{64}{9} \zeta_{3}\rb \cf^2 
          - \lb\f{23480}{243}  - \f{352}{9} \zeta_{3}\rb \ca \cf \right. \nonumber\\ & &\left.
          - \lb\f{30998}{243} + \f{128}{3} \zeta_{3}\rb \ca^2 
          \right] 
           +\Nf \Ngl  \lb\f{704}{9} - \f{512}{3}  \zeta_{3}\rb \dQAfNg \nonumber\\ & &
          - \Ngl^2 \Nf \ca^2 \tr \left[
            \f{308}{243} \cf
          + \f{934}{243} \ca \right]
          -\Ngl\Nf^2\ca \tr^2 \left[
           \f{1232}{243} \cf
          + \f{1252}{243} \ca \right] \nonumber         
\eea
in agreement with \cite{BednyakovTalkBetags4l}.

\section{Conclusions \label{last}}

We have presented an analytical result for the four-loop $\beta$-function of the strong coupling $\gs$ in a QCD-like model
with arbitrarily many fermion representations. As an application we have given the result for the four-loop $\beta$-function of
QCD plus gluinos.

\section*{Acknowledgements}
I would like to thank K.~G.~Chetyrkin for valuable discussions and useful comments on this paper.
This research was supported in part by the Swiss National Science Foundation (SNF) under contract BSCGI0\_157722.

\bibliographystyle{JHEP}

\bibliography{LiteraturSM}

\providecommand{\href}[2]{#2}\begingroup\raggedright\begin{thebibliography}{10}

\bibitem{Mikhailov:2004iq}
S.~V. Mikhailov, \emph{{Generalization of BLM procedure and its scales in any
  order of pQCD: A Practical approach}},
  \href{http://dx.doi.org/10.1088/1126-6708/2007/06/009}{\emph{JHEP} {\bf 06}
  (2007) 009}, [\href{http://arxiv.org/abs/hep-ph/0411397}{{\tt
  hep-ph/0411397}}].

\bibitem{Kataev:2014zwa}
A.~L. Kataev, \emph{{The generalized BLM approach to fix scale-dependence in
  QCD: the current status of investigations}},
  \href{http://dx.doi.org/10.1088/1742-6596/608/1/012078}{\emph{J. Phys. Conf.
  Ser.} {\bf 608} (2015) 012078}, [\href{http://arxiv.org/abs/1411.2257}{{\tt
  1411.2257}}].

\bibitem{Kataev:2014jba}
A.~L. Kataev and S.~V. Mikhailov, \emph{{Generalization of the
  Brodsky-Lepage-Mackenzie optimization within the beta-expansion and the
  principle of maximal conformality}},
  \href{http://dx.doi.org/10.1103/PhysRevD.91.014007}{\emph{Phys. Rev.} {\bf
  D91} (2015) 014007}, [\href{http://arxiv.org/abs/1408.0122}{{\tt
  1408.0122}}].

\bibitem{Brodsky:2013vpa}
S.~J. Brodsky, M.~Mojaza and X.-G. Wu, \emph{{Systematic Scale-Setting to All
  Orders: The Principle of Maximum Conformality and Commensurate Scale
  Relations}}, \href{http://dx.doi.org/10.1103/PhysRevD.89.014027}{\emph{Phys.
  Rev.} {\bf D89} (2014) 014027}, [\href{http://arxiv.org/abs/1304.4631}{{\tt
  1304.4631}}].

\bibitem{PhysRevLett.30.1343}
D.~J. Gross and F.~Wilczek, \emph{{Ultraviolet Behavior of Non-Abelian Gauge
  Theories}}, \href{http://dx.doi.org/10.1103/PhysRevLett.30.1343}{\emph{Phys.
  Rev. Lett.} {\bf 30} (1973) 1343--1346}.

\bibitem{PhysRevLett.30.1346}
H.~D. Politzer, \emph{{Reliable Perturbative Results for Strong
  Interactions?}},
  \href{http://dx.doi.org/10.1103/PhysRevLett.30.1346}{\emph{Phys. Rev. Lett.}
  {\bf 30} (1973) 1346--1349}.

\bibitem{Jones1974531}
D.~Jones, \emph{Two-loop diagrams in yang-mills theory},
  \href{http://dx.doi.org/10.1016/0550-3213(74)90093-5}{\emph{Nuclear Physics
  B} {\bf 75} (1974) 531--538}.

\bibitem{Tarasov:1976ef}
O.~Tarasov and A.~Vladimirov, \emph{{Two Loop Renormalization of the Yang-Mills
  Theory in an Arbitrary Gauge}}, {\emph{Sov.J.Nucl.Phys.} {\bf 25} (1977)
  585}.

\bibitem{PhysRevLett.33.244}
W.~E. Caswell, \emph{{Asymptotic Behavior of Non-Abelian Gauge Theories to
  Two-Loop Order}},
  \href{http://dx.doi.org/10.1103/PhysRevLett.33.244}{\emph{Phys.Rev.Lett.}
  {\bf 33} (1974) 244--246}.

\bibitem{Egorian:1978zx}
E.~Egorian and O.~Tarasov, \emph{{Two loop renormalization of the QCD in an
  arbitrary gauge}}, {\emph{Teor.Mat.Fiz.} {\bf 41} (1979) 26--32}.

\bibitem{Tarasov:1980au}
O.~Tarasov, A.~Vladimirov and A.~Y. Zharkov, \emph{{The Gell-Mann-Low Function
  of QCD in the Three Loop Approximation}},
  \href{http://dx.doi.org/10.1016/0370-2693(80)90358-5}{\emph{Phys.Lett.} {\bf
  B93} (1980) 429--432}.

\bibitem{3loopbetaqcd}
S.~Larin and J.~Vermaseren, \emph{{The Three loop QCD Beta function and
  anomalous dimensions}},
  \href{http://dx.doi.org/10.1016/0370-2693(93)91441-O}{\emph{Phys. Lett.} {\bf
  B303} (1993) 334--336}, [\href{http://arxiv.org/abs/hep-ph/9302208}{{\tt
  hep-ph/9302208}}].

\bibitem{4loopbetaqcd}
T.~van Ritbergen, J.~Vermaseren and S.~Larin, \emph{{The Four loop beta
  function in quantum chromodynamics}},
  \href{http://dx.doi.org/10.1016/S0370-2693(97)00370-5}{\emph{Phys. Lett.}
  {\bf B400} (1997) 379--384}, [\href{http://arxiv.org/abs/hep-ph/9701390}{{\tt
  hep-ph/9701390}}].

\bibitem{Czakon:2004bu}
M.~Czakon, \emph{{The Four-loop QCD beta-function and anomalous dimensions}},
  \href{http://dx.doi.org/10.1016/j.nuclphysb.2005.01.012}{\emph{Nucl.Phys.}
  {\bf B710} (2005) 485--498}, [\href{http://arxiv.org/abs/hep-ph/0411261}{{\tt
  hep-ph/0411261}}].

\bibitem{PhysRevLett.108.151602}
L.~N. Mihaila, J.~Salomon and M.~Steinhauser, \emph{{Gauge coupling beta
  functions in the standard model to three loops}},
  \href{http://dx.doi.org/10.1103/PhysRevLett.108.151602}{\emph{Phys. Rev.
  Lett.} {\bf 108} (2012) 151602}.

\bibitem{Mihaila:2012pz}
L.~N. Mihaila, J.~Salomon and M.~Steinhauser, \emph{{Renormalization constants
  and beta functions for the gauge couplings of the Standard Model to
  three-loop order}},
  \href{http://dx.doi.org/10.1103/PhysRevD.86.096008}{\emph{Phys. Rev. D} {\bf
  86} (2012) 096008}, [\href{http://arxiv.org/abs/1208.3357}{{\tt 1208.3357}}].

\bibitem{Bednyakov:2012rb}
A.~Bednyakov, A.~Pikelner and V.~Velizhanin, \emph{{Anomalous dimensions of
  gauge fields and gauge coupling beta-functions in the Standard Model at three
  loops}}, \href{http://dx.doi.org/10.1007/JHEP01(2013)017}{\emph{JHEP} {\bf
  1301} (2013) 017}, [\href{http://arxiv.org/abs/1210.6873}{{\tt 1210.6873}}].

\bibitem{Chetyrkin:2012rz}
K.~Chetyrkin and M.~Zoller, \emph{{Three-loop $\beta$-functions for top-Yukawa
  and the Higgs self-interaction in the Standard Model}},
  \href{http://dx.doi.org/10.1007/JHEP06(2012)033}{\emph{JHEP} {\bf 1206}
  (2012) 033}, [\href{http://arxiv.org/abs/1205.2892}{{\tt 1205.2892}}].

\bibitem{Bednyakov:2012en}
A.~Bednyakov, A.~Pikelner and V.~Velizhanin, \emph{{Yukawa coupling
  beta-functions in the Standard Model at three loops}},
  \href{http://dx.doi.org/10.1016/j.physletb.2013.04.038}{\emph{Phys.Lett.}
  {\bf B722} (2013) 336--340}, [\href{http://arxiv.org/abs/1212.6829}{{\tt
  1212.6829}}].

\bibitem{Chetyrkin:2013wya}
K.~G. Chetyrkin and M.~F. Zoller, \emph{{$\beta$-function for the Higgs
  self-interaction in the Standard Model at three-loop level}},
  \href{http://dx.doi.org/10.1007/JHEP04(2013)091}{\emph{JHEP} {\bf 04} (2013)
  091}, [\href{http://arxiv.org/abs/1303.2890}{{\tt 1303.2890}}].

\bibitem{Bednyakov:2013eba}
A.~Bednyakov, A.~Pikelner and V.~Velizhanin, \emph{{Higgs self-coupling
  beta-function in the Standard Model at three loops}},
  \href{http://dx.doi.org/10.1016/j.nuclphysb.2013.07.015}{\emph{Nucl.Phys.}
  {\bf B875} (2013) 552--565}, [\href{http://arxiv.org/abs/1303.4364}{{\tt
  1303.4364}}].

\bibitem{Bednyakov:2013cpa}
A.~V. Bednyakov, A.~F. Pikelner and V.~N. Velizhanin, \emph{{Three-loop Higgs
  self-coupling beta-function in the Standard Model with complex Yukawa
  matrices}},
  \href{http://dx.doi.org/10.1016/j.nuclphysb.2013.12.012}{\emph{Nucl. Phys.}
  {\bf B879} (2014) 256--267}, [\href{http://arxiv.org/abs/1310.3806}{{\tt
  1310.3806}}].

\bibitem{Bednyakov:2015ooa}
A.~V. Bednyakov and A.~F. Pikelner, \emph{{Four-loop strong coupling
  beta-function in the Standard Model}},
  \href{http://arxiv.org/abs/1508.02680}{{\tt 1508.02680}}.

\bibitem{Zoller:2015tha}
M.~F. Zoller, \emph{{Top-Yukawa effects on the $\beta$-function of the strong
  coupling in the SM at four-loop level}},
  \href{http://dx.doi.org/10.1007/JHEP02(2016)095}{\emph{JHEP} {\bf 02} (2016)
  095}, [\href{http://arxiv.org/abs/1508.03624}{{\tt 1508.03624}}].

\bibitem{Martin:2015eia}
S.~P. Martin, \emph{{Four-loop Standard Model effective potential at leading
  order in QCD}},
  \href{http://dx.doi.org/10.1103/PhysRevD.92.054029}{\emph{Phys. Rev.} {\bf
  D92} (2015) 054029}, [\href{http://arxiv.org/abs/1508.00912}{{\tt
  1508.00912}}].

\bibitem{Chetyrkin:2016ruf}
K.~G. Chetyrkin and M.~F. Zoller, \emph{{Leading QCD-induced four-loop
  contributions to the $\beta$-function of the Higgs self-coupling in the SM
  and vacuum stability}},
  \href{http://dx.doi.org/10.1007/JHEP06(2016)175}{\emph{JHEP} {\bf 06} (2016)
  175}, [\href{http://arxiv.org/abs/1604.00853}{{\tt 1604.00853}}].

\bibitem{Baikov:2016tgj}
P.~A. Baikov, K.~G. Chetyrkin and J.~H. K{\"uhn}, \emph{{Five-Loop Running of
  the QCD coupling constant}},  \href{http://arxiv.org/abs/1606.08659}{{\tt
  1606.08659}}.

\bibitem{Luthe:2016ima}
T.~Luthe, A.~Maier, P.~Marquard and Y.~Schroder, \emph{{Towards the five-loop
  Beta function for a general gauge group}},
  \href{http://arxiv.org/abs/1606.08662}{{\tt 1606.08662}}.

\bibitem{Jones:1974pg}
D.~R.~T. Jones, \emph{{Asymptotic Behavior of Supersymmetric Yang-Mills
  Theories in the Two Loop Approximation}},
  \href{http://dx.doi.org/10.1016/0550-3213(75)90256-4}{\emph{Nucl. Phys.} {\bf
  B87} (1975) 127}.

\bibitem{Clavelli:1996pz}
L.~Clavelli, P.~W. Coulter and L.~R. Surguladze, \emph{{Gluino contribution to
  the three loop beta function in the minimal supersymmetric standard model}},
  \href{http://dx.doi.org/10.1103/PhysRevD.55.4268}{\emph{Phys. Rev.} {\bf D55}
  (1997) 4268--4272}, [\href{http://arxiv.org/abs/hep-ph/9611355}{{\tt
  hep-ph/9611355}}].

\bibitem{BednyakovTalkBetags4l}
A.~V. Bednyakov and A.~F. Pikelner, \emph{{On the four-loop strong coupling
  beta function in the SM}},  presented at QUARKS-2016, Russia.

\bibitem{QGRAF}
P.~Nogueira, \emph{{Automatic Feynman graph generation}},
  \href{http://dx.doi.org/10.1006/jcph.1993.1074}{\emph{J. Comput. Phys.} {\bf
  105} (1993) 279--289}.

\bibitem{Seidensticker:1999bb}
T.~Seidensticker, \emph{{Automatic application of successive asymptotic
  expansions of Feynman diagrams}},  in \emph{{6th International Workshop on
  New Computing Techniques in Physics Research: Software Engineering,
  Artificial Intelligence Neural Nets, Genetic Algorithms, Symbolic Algebra,
  Automatic Calculation (AIHENP 99) Heraklion, Crete, Greece, April 12-16,
  1999}}, 1999.
\newblock \href{http://arxiv.org/abs/hep-ph/9905298}{{\tt hep-ph/9905298}}.

\bibitem{Harlander:1997zb}
R.~Harlander, T.~Seidensticker and M.~Steinhauser, \emph{{Complete corrections
  of Order alpha alpha-s to the decay of the Z boson into bottom quarks}},
  \href{http://dx.doi.org/10.1016/S0370-2693(98)00220-2}{\emph{Phys.Lett.} {\bf
  B426} (1998) 125--132}, [\href{http://arxiv.org/abs/hep-ph/9712228}{{\tt
  hep-ph/9712228}}].

\bibitem{Zoller:2014xoa}
M.~Zoller, \emph{{Three-loop beta function for the Higgs self-coupling}},
  {\emph{PoS} {\bf LL2014} (2014) 014},
  [\href{http://arxiv.org/abs/1407.6608}{{\tt 1407.6608}}].

\bibitem{Misiak:1994zw}
M.~Misiak and M.~M{\"u}nz, \emph{{Two loop mixing of dimension five flavor
  changing operators}},
  \href{http://dx.doi.org/10.1016/0370-2693(94)01553-O}{\emph{Phys. Lett.} {\bf
  B344} (1995) 308--318}, [\href{http://arxiv.org/abs/hep-ph/9409454}{{\tt
  hep-ph/9409454}}].

\bibitem{beta_den_comp}
K.~G. Chetyrkin, M.~Misiak and M.~M{\"u}nz, \emph{{Beta functions and anomalous
  dimensions up to three loops}},
  \href{http://dx.doi.org/10.1016/S0550-3213(98)00122-9}{\emph{Nucl. Phys.}
  {\bf B518} (1998) 473--494}, [\href{http://arxiv.org/abs/hep-ph/9711266}{{\tt
  hep-ph/9711266}}].

\bibitem{Vermaseren:2000nd}
J.~A.~M. Vermaseren, \emph{{New features of FORM}},
  \href{http://arxiv.org/abs/math-ph/0010025}{{\tt math-ph/0010025}}.

\bibitem{Tentyukov:2007mu}
M.~Tentyukov and J.~A.~M. Vermaseren, \emph{{The Multithreaded version of
  FORM}}, \href{http://dx.doi.org/10.1016/j.cpc.2010.04.009}{\emph{Comput.
  Phys. Commun.} {\bf 181} (2010) 1419--1427},
  [\href{http://arxiv.org/abs/hep-ph/0702279}{{\tt hep-ph/0702279}}].

\bibitem{MATAD}
M.~Steinhauser, \emph{{MATAD: A program package for the computation of massive
  tadpoles}},
  \href{http://dx.doi.org/10.1016/S0010-4655(00)00204-6}{\emph{Comput. Phys.
  Commun.} {\bf 134} (2001) 335--364},
  [\href{http://arxiv.org/abs/hep-ph/0009029}{{\tt hep-ph/0009029}}].

\bibitem{Smirnov:2008iw}
A.~Smirnov, \emph{{Algorithm FIRE -- Feynman Integral REduction}},
  \href{http://dx.doi.org/10.1088/1126-6708/2008/10/107}{\emph{JHEP} {\bf 0810}
  (2008) 107}, [\href{http://arxiv.org/abs/0807.3243}{{\tt 0807.3243}}].

\bibitem{Smirnov:2014hma}
A.~V. Smirnov, \emph{{FIRE5: a C++ implementation of Feynman Integral
  REduction}},
  \href{http://dx.doi.org/10.1016/j.cpc.2014.11.024}{\emph{Comput.Phys.Commun.}
  {\bf 189} (2014) 182--191}, [\href{http://arxiv.org/abs/1408.2372}{{\tt
  1408.2372}}].

\bibitem{COLOR}
T.~Van~Ritbergen, A.~Schellekens and J.~Vermaseren, \emph{Group theory factors
  for feynman diagrams}, {\emph{International Journal of Modern Physics A} {\bf
  14} (1999) 41--96}.

\end{thebibliography}\endgroup

\end{document}